%% file: main.tex
\documentclass[%
amsmath, amssymb, aps, twocolumn,
pra, superscriptaddress
]{revtex4-2}
\input{preamble.tex}
\RequirePackage[T1]{fontenc}
\RequirePackage[utf8]{inputenc}
\RequirePackage{lmodern}
\usepackage{xcolor}

\pdfoutput=1

\input{glossary}
\newcommand{\rate}[2]{\ensuremath{\Gamma_{#1,{\rm #2}}}}
\newcommand{\tsample}{\ensuremath{T_{\rm s}}}
\begin{document}

\title{Beating the thermal limit of qubit initialization with a Bayesian Maxwell's demon}

\author{Mark A. I. Johnson}
\affiliation{School of Electrical Engineering and Telecommunications, UNSW Sydney, Sydney, New South Wales 2052, Australia}
\affiliation{Centre of Excellence for Quantum Computation \& Communication Technology, Australia}

\author{Mateusz T. M\k{a}dzik}
\affiliation{School of Electrical Engineering and Telecommunications, UNSW Sydney, Sydney, New South Wales 2052, Australia}
\affiliation{Centre of Excellence for Quantum Computation \& Communication Technology, Australia}

\author{Fay E. Hudson}
\affiliation{School of Electrical Engineering and Telecommunications, UNSW Sydney, Sydney, New South Wales 2052, Australia}

\author{Kohei M. Itoh}
\affiliation{School of Fundamental Science and Technology, Keio University, Kohoku-ku, Yokohama,  Kanagawa 223-8522, Japan}

\author{Alexander M. Jakob}
\affiliation{School of Physics, University of Melbourne, Melbourne, Victoria 3010, Australia}
\affiliation{Centre of Excellence for Quantum Computation \& Communication Technology, Australia}

\author{David N. Jamieson}
\affiliation{School of Physics, University of Melbourne, Melbourne, Victoria 3010, Australia}
\affiliation{Centre of Excellence for Quantum Computation \& Communication Technology, Australia}

\author{Andrew Dzurak}
\affiliation{School of Electrical Engineering and Telecommunications, UNSW Sydney, Sydney, New South Wales 2052, Australia}

\author{Andrea Morello}%
\email{a.morello@unsw.edu.au}
\affiliation{School of Electrical Engineering and Telecommunications, UNSW Sydney, Sydney, New South Wales 2052, Australia}
\affiliation{Centre of Excellence for Quantum Computation \& Communication Technology, Australia}%

\date{\today}
\begin{abstract}
    Fault-tolerant quantum computing requires initializing the quantum register in a well-defined fiducial state.
    In solid-state systems, this is typically achieved through thermalization to a cold reservoir, such that the initialization fidelity is fundamentally limited by temperature. 
    Here, we present a method of preparing a fiducial quantum state with a confidence beyond the thermal limit. 
    It is based on real time monitoring of the qubit through a negative-result measurement -- the equivalent of a `Maxwell's demon' that triggers the experiment only upon the appearance of a qubit in the lowest-energy state. 
    We experimentally apply it to initialize an electron spin qubit in silicon, achieving a ground-state initialization fidelity of 98.9(4)\%, corresponding to a 20$
    \times$ reduction in initialization error compared to the unmonitored system. A fidelity approaching 99.9\% could be achieved with realistic improvements in the bandwidth of the amplifier chain or by slowing down the rate of electron tunneling from the reservoir.
    We use a nuclear spin ancilla, measured in quantum nondemolition mode, to prove the value of the electron initialization fidelity far beyond the intrinsic fidelity of the electron readout. 
    However, the method itself does not require an ancilla for its execution, saving the need for additional resources. 
    The quantitative analysis of the initialization fidelity reveals that a simple picture of spin-dependent electron tunneling does not correctly describe the data.
    Our digital `Maxwell's demon' can be applied to a wide range of quantum systems, with minimal demands on control and detection hardware.  
\end{abstract}

\maketitle

\section{Introduction}

Fault-tolerant quantum computing places strict requirements on preparation, control and readout of qubits. Much emphasis is placed in the literature on the fidelity of quantum gate operations, but the most effective error correction codes demand an abundance of initialization and measurement steps. A well-known example is the surface code: in this scheme, fault-tolerant operation of a logical qubit is achieved when the probability of an error occurring on one- and two-qubit gates, \emph{and} initialization and measurement, are \emph{all} below 0.56\% \cite{Fowler2012}. Similarly, a fault-tolerance error threshold of 4.8\% is obtained for topological color codes, assuming both qubit errors and \gls*{spam} errors have the same rate \cite{Andrist2011}.

Many physical qubit platforms have now reached quantum logic gate errors close to or below some fault-tolerance thresholds \cite{Benhelm2008,Barends2014,abobeih2022fault}, including  nuclear \cite{madzik2022precision} and electron \cite{xue2022quantum,noiri2022fast,mills2022} spin qubits in silicon. 
Not as common, however, is the achievement of equivalently low \gls*{spam} errors. 
The lowest errors are typically achieved in systems that allow for continuous or repetitive \gls*{qnd} readout, such as nuclear spins \cite{Neumann2010,Pla2013} and superconducting transmon qubits \cite{Vijay2011,Jeffrey2014}. Projective \gls*{qnd} measurements combined with feedback can be used to initialize the qubits with high fidelity \cite{riste2012feedback,campagne2013persistent}.
Other qubit types, like superconducting phase qubits \cite{Katz2006} and electron spins in semiconductors \cite{Elzerman2004,Morello2010}, adopt instead a measurement method based upon energy-dependent tunneling. 
There, the high-energy state of the qubit is detected through its much higher probability of tunneling out of the confining potential barrier; the tunneling event, in turn, causes a change in the state of a nearby sensor. 
This method naturally results in a \gls*{spam} fidelity limited by the thermal population of the energy eigenstates. 
Recent work on electron spin qubits in Si/SiGe quantum dots \cite{Yoneda2020,Xue2020} has begun to address the issue of SPAM errors by implementing \gls*{qnd} readout via a quantum logic operation with an ancilla dot.

In this paper we describe and demonstrate a method to beat the thermal limit for initializing the ground state of an electron spin qubit, which is read out via energy-dependent tunneling. 
Conceptually, energy-dependent tunneling is a form of negative-result measurement \cite{Dicke1981}. The theory of negative-result measurements has been extended to the case of a driven qubit \cite{Ruskov2007} at zero temperature. Here, we focus instead on the finite-temperature case in the absence of drive, by extending the model originally presented by Ruskov et al. \cite{Ruskov2007} to account for thermally induced tunneling out of the ground state, which represents a measurement error. The goal of our work is to reduce such error far below the thermal limit.
We use the spin of the electron bound to an ion-implanted $^{31}$P donor in silicon as our qubit \cite{Pla2012}, which naturally comes equipped with a spin-$1/2$ nucleus coupled to it. 
We use the nucleus as an ancilla system that can be read out with near-unity fidelity \cite{Pla2013,Dehollain2016}, solely to prove the effectiveness of our electron initialization method.
The method itself does not require an ancilla to operate and, thus, does not demand additional quantum resources.

The electron spin qubit is initialized by a real-time monitoring system, implemented in \gls*{fpga} hardware, which acts as a Bayesian `Maxwell's demon' by selectively preparing the lowest-energy spin state of an electron, which is initially drawn from a charge reservoir at nonzero temperature.
In the original thought experiment, the demon separates high-energy from low-energy particles in a gas at no energy cost, and the separated gas is then operated as a heat engine to extract work from the system.
This appears to violate the second law of thermodynamics; however, an information theoretic approach reveals the entropic cost of the demon resetting its knowledge at the beginning of each cycle\cite{Bennett1973,Landauer1961}.
Maxwell's demon has recently been applied in charge\cite{Averin2011,Schaller2011,Koski2014} and superconducting\cite{Cottet2017,Masuyama2018} qubit systems to study the thermodynamic cost of quantum information processing and to extract work.
Here we adopt the Maxwell's demon idea to draw an electron from a warm charge reservoir.
By monitoring in real-time the charge state of the donor, our demon is able to obtain a higher probability of the electron being in the ground state than would be achievable without the demon's careful observation.

Our device shares many similarities with the seminal experiment of Koski \emph{et al.} \cite{Koski2014}, who realize a Szilard engine by using a single-electron box, whose charge state is monitored in real time by a single-electron transistor embodying the Maxwell's demon. In their work, the information acquired by the demon is used to then extract work equal to $k_{\rm B}T \ln{2}$ from a thermal bath. However, our work differs from previous Maxwell's demon experiments in that our interest is not in the thermodynamics itself, but in the impact of the information gain on the initialization fidelity of a qubit. Furthermore, because of the quantum information focus of our work, the initialization speed is of paramount importance. Our readout system is, thus, 3 orders of magnitude faster than the one used by Koski \emph{et al.} \cite{Koski2014} in their Szilard engine experiment.

\section{Electron spin initialization and readout}

We consider electron spin qubits in semiconductors \cite{Chatterjee2021}. 
An external magnetic field, $B_0$, creates an energy splitting $\Delta E_{\rm Z} = E_\uparrow -E_\downarrow = h\gamma_{\rm e} B_0$, between the spin-down ($\ket{\downarrow}$) and spin-up ($\ket{\uparrow}$) states due to the Zeeman effect, where $\gamma_{\rm e}$ is the electron gyromagnetic ratio and $h$ is the Planck constant.
For electrons in silicon, $\gamma_{\rm e} \approx 28$~GHz/T, such that $B_0 = 1$~T results in $\Delta E_{\rm Z}/k_{\rm B} \approx 1.34$~K, where $k_{\rm B}$ is the Boltzmann constant. 
From a microwave engineering point of view, it is costly and challenging to operate spin qubits at frequencies above 40~GHz, which means that $\Delta E_{\rm Z} / k_{\rm B}$ is limited to approximately 2~K for practical purposes, which corresponds to $B_0 \approx 1.4$~T in silicon.

A natural way to initialize such a spin qubit is to cool down the host material to a temperature $T \ll \Delta E_{\rm Z} / k_{\rm B}$ and wait a time approximately $5 T_{\rm 1e}$, where $T_{\rm 1e}$ is the electron spin-lattice relaxation time. 
At low magnetic fields and low temperatures (typically about $100$~mK), $T_{\rm 1e}$ usually exceeds one second \cite{Morello2010,Yang2013,Watson2017,Camenzind2018,borjans2019,Tenberg2019}. 
Therefore, this method is practical only if the electron spin possesses an energy level structure which results in relaxation '`hot spots'' \cite{Stano2006}, where the Zeeman splitting crosses an orbital or valley splitting, shortening $T_{\rm 1e}$ to the microsecond range. 
Hot spots exist in quantum dots \cite{Yang2013,borjans2019} but not in donors \cite{Watson2017,Tenberg2019}.

\begin{figure}[ht]
    \includegraphics[
        width=\linewidth]{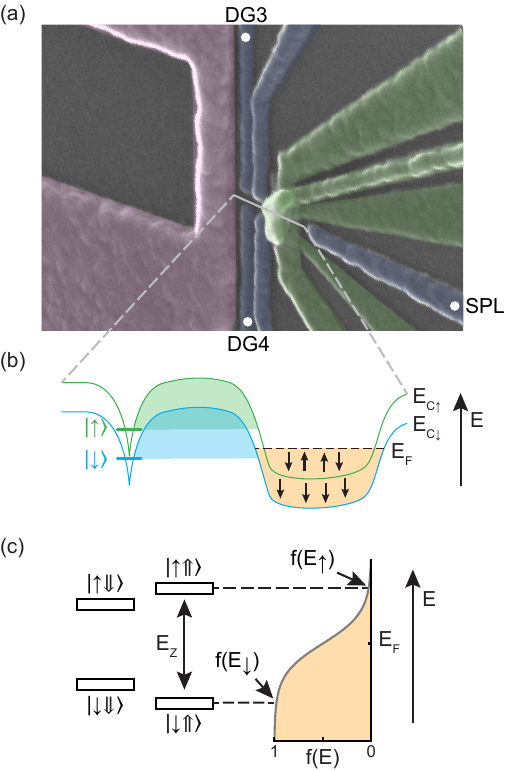}
    \caption{
    (a) False-colored micrograph of a device from the same batch as the one used in this paper.
    The \acrshort*{set} for electron spin readout and initialization is shown on the right in green.
    The microwave antenna, used for \acrshort*{esr} and \acrshort*{nmr}, is shown on the left in pink.
    Pulsing gates DG3, DG4, and SPL (identified by white dots) form a virtual gate VG.
    (b) Schematic energy landscape of the device. 
    The donor binding potential confines an electron in the spin-down (blue line) or spin-up (green line) state. 
    The conduction band edge is indicated for both spin-up ($E_{\rm c\uparrow}$) and spin-down ($E_{\rm c\downarrow}$) electrons, split by the Zeeman energy $E_{\rm Z}$.
    The green and blue shaded regions illustrate that $\ket{\uparrow}$ and $\ket{\downarrow}$ electrons are expected to face the same energy barrier to the reservoir upon elastic tunneling (see Sec.~\ref{sec:electron_temp}).
    The electron is initially drawn from the \acrshort*{set} island, shown in tan, which acts as a finite-temperature charge reservoir.
    (c) Level diagram of the electron-nuclear spin system (not to scale), alongside the occupied states in the \gls*{set} island, which follow the Fermi-Dirac distribution $f(E)$.
    }
    \label{fig:main_figure_1}
\end{figure}

More commonly, the electron spin is initialized by the same method used for readout. 
Here we focus on the case where the readout occurs via energy-dependent tunneling to a Fermi reservoir \cite{Elzerman2004,Morello2010,Watson2015}. We use spin-to-charge conversion to measure the state of an electron spin bound to a $^{31}$P donor in isotopically enriched $^{28}$Si \cite{Morello2020}. The device is placed in a magnetic field $B_0 = \SI{1.423}{\tesla}$ and cooled to $T\ll 1$~K with a dilution refrigerator.
The donor electron is tunnel coupled to a nearby \gls*{set} which is tuned to be in Coulomb blockade (no current flowing) when the donor is in the neutral charge state ($D^0$). 
The blockade is lifted when the donor becomes ionized and positively charged ($D^+$) \cite{Morello2009} upon tunneling of its electron onto the \gls*{set}.
The \gls*{set} island can be approximated as a Fermi reservoir, as shown in Figure \ref{fig:main_figure_1} (this approximation is examined in Sec.~\ref{sec:electron_temp}). The fraction of occupied electron states, $f$, follows the Fermi function
\begin{equation}
    f(E) = \left[1 + \exp\left(\dfrac{E- E_{\rm F}}{k_B T_{\rm e}}\right)\right]^{-1},
    \label{eq:fermi_function}
\end{equation}
where $E_{\rm F}$ is the Fermi energy.

When $\Delta E_{\rm Z} \gg k_B T_{\rm e}$, and the electrochemical potential of the donor is tuned about the Fermi level, a $\ket{\uparrow}$ electron can tunnel to the reservoir at a rate \rate{\uparrow}{out}, which is high because of the large density of empty states in the reservoir, i.e. $1-f(E_{\uparrow}) \approx 1$. 
Tunneling of a $\ket{\downarrow}$ electron, at a rate \rate{\downarrow}{out}, is greatly suppressed by the lack of available states, $1-f(E_{\downarrow}) \ll 1$. 

Single-shot electron spin readout is performed by observing the \gls*{set} current for a duration $t_{\rm read}$. 
Traces where the current shows a step increase indicate an ionization event (transitioning from $D^0$ to $D^+$). 
The ability to distinguish $\ket{\uparrow}$ and $\ket{\downarrow}$ tunneling events is given by the ratio of the tunnel-out rates for the two orientations.
Because $\rate{\uparrow}{out} \gg \rate{\downarrow}{out}$, we identify this event with tunneling out of a $\ket{\uparrow}$ electron.
Conversely, the absence of current steps is identified with a $\ket{\downarrow}$ electron remaining on the donor. 
In a broader quantum measurement context, this type of qubit readout belongs to the class of negative-result measurements \cite{Dicke1981}: Even the absence of an event (in this case, electron tunneling) provides information on the state of the monitored quantum system. This should not be confused with whether the sensor output that signals a tunneling event is a positive or a negative current (or voltage) \cite{ciriano2021spin}. Here the ``negative result'' is the \emph{absence} of a sensor signal, regardless of its sign.

An alternative method to read out an electron spin state is Pauli spin blockade (PSB). PSB distinguishes the triplet states of two spins from the singlet state $\ket{S} = \ket{\uparrow\downarrow} - \ket{\downarrow\uparrow}$, as only a spin singlet can occupy the same spatial quantum state [e.g. a (2, 0) charge configuration of two dots \cite{Petta2005}]. PSB spin mapping errors arise from triplet-to-singlet relaxation during the readout window, resulting in a tunneling event for an initially blockaded spin configuration. As a result of PSB readout, the final two-spin state is either a singlet in the (2, 0) charge configuration or one of the triplet states in the (1, 1) charge configuration. Under certain conditions, PSB can be used instead to discern the two-electron parity, i.e. discriminate $\ket{\downarrow\uparrow},\ket{\uparrow\downarrow}$ from $\ket{\downarrow\downarrow},\ket{\uparrow\uparrow}$ \cite{seedhouse2021pauli}.

This alone does not initialize a \emph{single} electron spin state. However, if one of the electrons can be independently initialized to the fiducial state $\ket{\downarrow}$ through e.g. energy-selective tunneling or $T_1$ relaxation, then PSB readout resulting in a triplet or even-parity state prepares the state $\ket{T_{-}} \equiv \ket{\downarrow\downarrow}$.

\section{Bayesian Maxwell's demon}
\label{sec:bayesian_maxwell_demon}

\begin{figure}[tp]
    \includegraphics{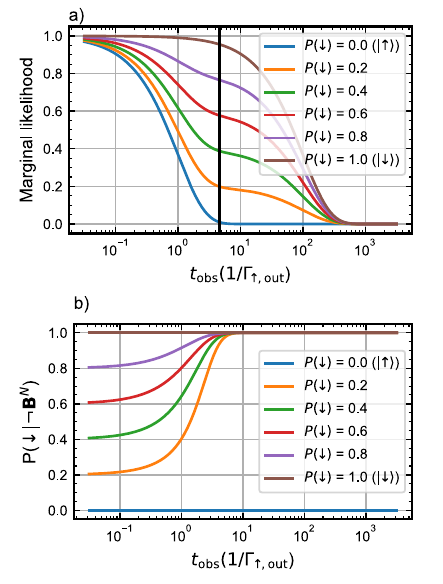}
    \caption{
    a) An example marginal likelihood function, $\likelihood{\downarrow}{\neg \bm{B}^N} \prob{\downarrow} + \likelihood{\uparrow}{\neg \bm{B}^N} \prob{\uparrow}$, (denominator of Eq.~\ref{eq:bayesian_update_full}), calculated assuming $\rate{\uparrow}{out} = 100 \rate{\downarrow}{out}$, as a function of the observation time $t_{\rm obs}$ expressed in units of the spin-up tunnel-out time $1/\rate{\uparrow}{out}$.
    For short $t_{\rm obs}$ the electron is unlikely to have tunneled from the donor.
    For long $t_{\rm obs}$, the electron escapes the donor regardless of its spin state. The ideal $t_{\rm obs}$ (heavy black line) for single-shot spin readout is found where the likelihood function exhibits a large contrast between its values for $P(\downarrow)=0$ and $P(\downarrow)=1$ (black line). For these tunnel rates, the ideal spin readout contrast is approximately 95\%.
    b) Bayesian update of the $\ket{\downarrow}$ probability.
    For increasing observation times where a tunneling event is not observed, the postmeasurement $\ket{\downarrow}$ probability increases toward unity.
    Different priors are shown to indicate that the prior spin population has only a small role in determining the final $\ket{\downarrow}$ probability after a sufficiently long observation time.
    With a prior probability of $P(\downarrow) = 0$, we know with certainty that the electron spin state is initially $\ket{\uparrow}$. 
    Therefore, no amount of new information (data gathered through observation) can update the $\ket{\downarrow}$ probability.
    }
    \label{fig:main_figure_model}
\end{figure}

\begin{figure*}[tp]
    \includegraphics[
        ]{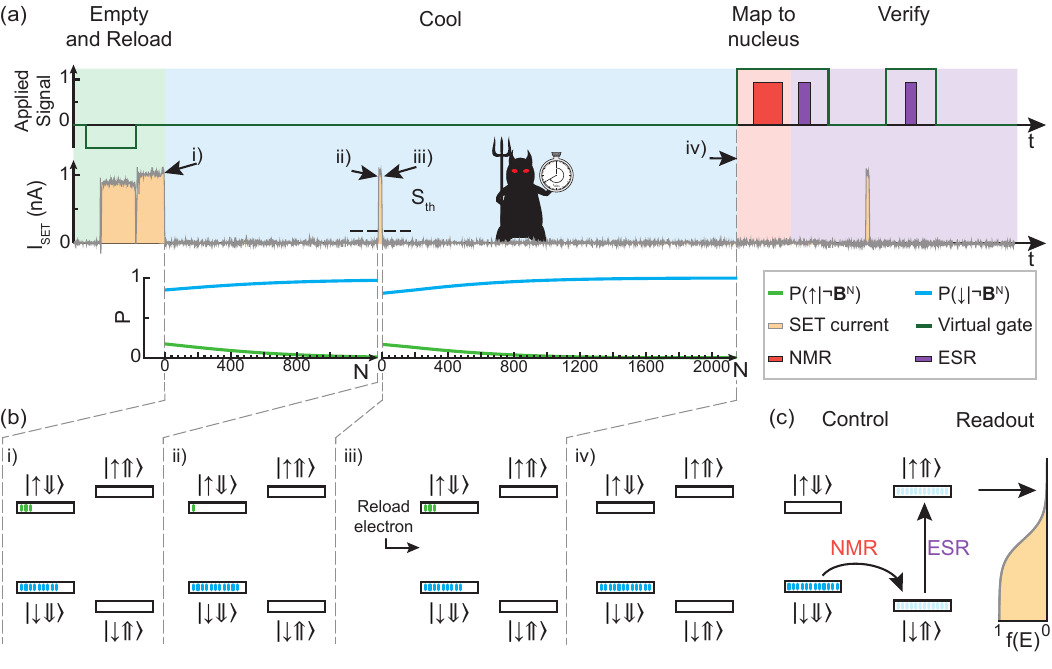}
    \caption{
    Electron spin initialization with Bayesian Maxwell's demon. 
    (a) A pulse on the virtual gate VG first empties the donor to remove the electron and then returns to the initialization position ($\mu_{\rm D} = E_{\rm F}$) to reload an electron from the \acrshort*{set} island. The electron loading is signalled by $I_{\rm SET}$ returning to approximately $0$.
    A Maxwell's demon (in this case, the FPGA that processes the values of $I_{\rm SET}$) observes the digital trace with a stopwatch to count to $N\tsample = t_{\rm obs}$, resetting the stopwatch when a tunneling event occurs.
    After a time $t_{\rm obs}$ of measuring the electron not tunneling, the electron is labeled as $\ket{\downarrow}$ with high confidence.
    (b) The electron spin populations are shown at each point (i) -- (iv) from panel (a), indicated by individual ticks.
    The $\ket{\downarrow}$ population increases with increasing $t_{\rm obs}$, but the spin preparation is lost if the electron tunnels away from the donor. The demon automatically restarts the process, until a long enough stretch with no tunnel-out events is observed.
    (c) Finally, the electron spin preparation is mapped to a nuclear spin flip with an \acrshort*{nmr} pulse, and the nuclear spin is read out via repeated \acrshort*{esr}.
    The initial nuclear spin state [$\ket{\Downarrow}$) and the prior spin-down probability ($\prob{\downarrow} = 0.75$' are chosen for illustration only.
    }
    \label{fig:main_figure_3}
\end{figure*}
To understand how the electron spin state populations change under the Maxwell's demon observation, we adopt a Bayesian update framework, where the knowledge of the state populations is updated with each measurement sample. This framework naturally describes a digital, discrete-time measurement. In Appendix~\ref{app:weak_measurement} we describe the equivalent continuous-time process.

In a setup as described above, the current through the \gls*{set} charge sensor is used to determine whether an electron has tunneled to a reservoir. The current is then converted to a voltage through a transimpedance amplifier at room temperature, integrated over a short duration $\tsample \approx \SI{10}{\micro\second}$, and digitized to yield a digital sample $D_n$. Each sample yields information on whether the electron has tunneled or not, which correlates to the electron spin state. This information is processed by a \gls*{fpga}, which enables real time decision making based on the information extracted from the sensor.

We define a Boolean parameter $B$, to represent the sensor signal exceeding a threshold $S_{\rm th}$ above which a tunneling event is recognized, i.e. $B = (D > S_{\rm th})$.
The logical complement to $B$ is denoted $\neg B$ and is recorded when $D \leq S_{\rm th}$.

The sensor measurement has a backaction on the electron spin, even when no tunneling event takes place \cite{Muhonen2018}. This constitutes a variable-strength measurement, which is elaborated in Appendix \ref{app:weak_measurement}.
The initial estimate of the $\ket{\downarrow}$ probability $P(\downarrow)$ prior to the measurement can be taken as $P(\downarrow) = f(E_{\downarrow})$, since the electron is loaded from the \gls*{set} island Fermi reservoir.
As the measurement proceeds, we describe the probability of having a $\ket{\downarrow}$ electron on the donor by an iterative Bayesian update process.
If no blip is detected in a set of $N$ measured samples $\neg \bm{B}^N = (\neg B_N, \neg B_{N-1}, \dots, \neg B_1)$, i.e.~the signal threshold is not exceeded, the prior $\ket{\downarrow}$ probability $\prob{\downarrow}$ is updated based on the likelihoods $\likelihood{\downarrow}{\neg \bm{B}^N}$ and $\likelihood{\uparrow}{\neg \bm{B}^N}$ of a $\ket{\downarrow}$ or $\ket{\uparrow}$ electron \emph{not} tunneling (see Appendix \ref{app:bayesian} for derivations).
Gathering $N$ consecutive samples that all show no blip gives the final posterior probability [see Figure \ref{fig:main_figure_model}b] as: 
\begin{align}
        \pgiv{\downarrow}{\neg \bm{B}^N} &= \dfrac{\likelihood{\downarrow}{\neg \bm{B}^N} \prob{\downarrow}}{\likelihood{\downarrow}{\neg \bm{B}^N} \prob{\downarrow} + \likelihood{\uparrow}{\neg \bm{B}^N} \prob{\uparrow}} \label{eq:bayesian_update_full}
        \\
         &= \dfrac{1}{1 + \dfrac{\likelihood{\uparrow}{\neg \bm{B}^N} \prob{\uparrow}}{\likelihood{\downarrow}{\neg \bm{B}^N} \prob{\downarrow}}} \\
        & = \Big[1 + \dfrac{1 - \prob{\downarrow}}{\prob{\downarrow}} e^{-N \tsample(\rate{\uparrow}{out} - \rate{\downarrow}{out})}  \Big]^{-1}, \label{eq:bayesian_update_spin_down}
\end{align}
which tends to 1 as $N\rightarrow \infty$.

If a tunneling event is observed, i.e.~a positive readout result, the knowledge of the Maxwell's demon is reset. In this manner, the positive result is simply discarded and initialization recommences when an electron rejoins the donor.

The denominator in Eq.~(\ref{eq:bayesian_update_full}) is the marginal likelihood [example shown in Figure~\ref{fig:main_figure_model}a]. It represents the likelihood of an electron not tunneling out of the donor after $N$ observations, given the initial probabilities of occupying each spin state.
The likelihood tends to zero for long observation times, which indicates that long records of no tunneling become increasingly rare.

This is the sense in which our setup operates as a Bayesian Maxwell's demon: as $N$ increases and the memory fills up with more samples of $\neg B$, the demon becomes increasingly confident that the electron, initially drawn from the warm reservoir, is in the ground state. In that sense, the electron can be thought of as being in equilibrium with a much colder reservoir than the one it is originally drawn from. Once the desired confidence is reached, the demon can give a trigger signal to start further quantum operations on the qubit.
Below we discuss how to quantify such confidence.

\section{Qubit initialization fidelity}
The qubit initialization fidelity is the probability of correctly preparing a $\ket{\downarrow}$ electron.  The basic method of initialization by spin-dependent tunneling, i.e. the same process used for electron spin readout, yields simply $\mathcal{F}_{\rm I}(0) = P(\downarrow)=f(E_{\downarrow})$, limited by the thermal broadening of the Fermi distribution. The Maxwell's demon observation improves this fidelity as $\mathcal{F}_{\rm I}(N) = {\pgiv{\downarrow}{\neg \bm{B}^N}}$ [Eq.~\ref{eq:bayesian_update_spin_down}]. However, we have no direct way to verify such improvement if we resort to an electron spin measurement with fidelity limited by the same thermally broadened reservoir. This limit can be circumvented  by introducing an ancilla qubit which can be read out repetitively in a quantum nondemolition (QND) fashion. For a P donor system, such an ancilla is naturally provided by the $^{31}$P nuclear spin. After the Maxwell's demon cooling operation, we map the state of the electron spin onto the nucleus \cite{Morton2008,Freer2017} by a simple \gls*{nmr} $\pi$ pulse conditional on the electron $\ket{\downarrow}$ state, followed by repetitive nuclear spin readout \cite{Pla2013}.

The total experiment fidelity $\mathcal{F}$ is thus composed of the fidelities of the three stages: electron initialization ($\mathcal{F}_{\rm I}$), \gls*{nmr} control ($\mathcal{F}_{\rm C}$) to map the electron state onto the nucleus, and nuclear readout ($\mathcal{F}_{\rm R}$):
\begin{equation}
    \mathcal{F} = \mathcal{F_{\rm I}} \cdot \mathcal{F_{\rm C}} \cdot \mathcal{F_{\rm R}}.
    \label{eq:experiment_fidelity}
\end{equation}
Therefore, to extract $\mathcal{F}_{\rm I}$ from the experiment we need to independently quantify $\mathcal{F}_{\rm C}$ and $\mathcal{F}_{\rm R}$. 
This is detailed in Appendices \ref{app:control_fidelity} and \ref{app:readout_fidelity} where we find $\mathcal{F}_{\rm C} = 99.5(3)\%$ and $\mathcal{F}_{\rm R} = 99.99\%$.

We stress that the Maxwell's demon cooling method does not require an ancilla to be implemented. The only role of the ancilla qubit here is to verify the effectiveness of the cooling method, to a precision better than the intrinsic electron readout fidelity.

\begin{figure}[tp]
    \includegraphics{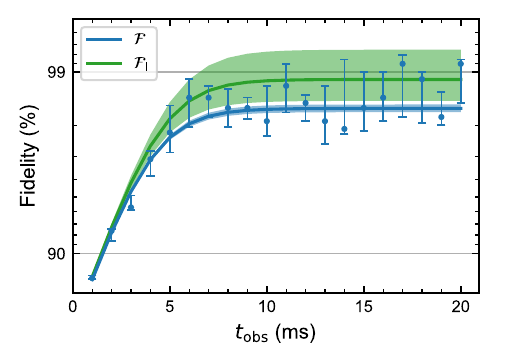}
    \caption{
    The $\ket{\downarrow}$ population increases with increasing observation time.
    The median experiment fidelity $\mathcal{F}$ is shown by the blue points with corresponding line of best fit to Eq.~\ref{eq:fit_function} (the blue shaded region indicates one standard deviation model uncertainty).
    The initialization fidelity $\mathcal{F}_{\rm I}$ is shown by the green line of best fit (one standard deviation model uncertainty indicated by the shaded green region).
    }
    \label{fig:main_figure_4}
\end{figure}

Figure \ref{fig:main_figure_4} shows the total experiment ($\mathcal{F}$) and initialization ($\mathcal{F}_{\rm I}$) fidelities of the donor electron spin, as a function of observation times $t_{\rm obs}$.
The error bars indicate the 25th and 75th percentiles of the data.
As the data are binomially distributed, they are highly skewed (non-normal) near zero.
Percentiles are shown instead of the standard deviation as one standard deviation above the mean may be greater than one, which is unphysical.

We observe that, in contrast with the ideal prediction of Eq.~\ref{eq:bayesian_update_spin_down}, the initialization fidelity $\mathcal{F}_{\rm I}$ asymptotically approaches an upper bound $1 - P_M$ for $P_M \neq 0$. 
We attribute this discrepancy to fast tunneling events, resulting in current blips that go undetected because they are shorter than the rise time $t_{\rm rise}$ of the transimpedance amplifier used to detect changes in the \gls*{set} current. Our amplifier has a low-pass cutoff frequency $f_{\rm c} = 50$~kHz, which yields a rise time (using a first-order approximation):
\begin{equation}
    t_{\rm rise} = -\frac{1}{2\pi f_{\rm c}} \log(1 - S_{\rm th}) \approx \SI{1.1}{\micro\second},
\end{equation}
where $S_{\rm th} = 0.3$ is the \gls*{set} signal threshold [Fig.~\ref{fig:main_figure_3}a] with the \gls*{set} signal rescaled between 0 and 1.
The tunnel-in rate $\rate{\uparrow}{in} + \rate{\downarrow}{in}$, which determines the average blip duration, is measured to be approximately $\SI{2700}{\second^{-1}}$ (Sec.~\ref{sec:electron_temp}).
The probability $P_M$ of missing a blip, i.e. the proportion of tunneling events from the reservoir to the donor that occur before the \gls*{set} signal rises to the signal threshold, is (see Appendix \ref{app:missing_blips} for details)
\begin{equation}
    P_M = 1 - \exp[-t_{\rm rise}(\rate{\uparrow}{in} + \rate{\downarrow}{in})] \approx 0.3\%. \label{eq:limit_infidelity_BW}
\end{equation}

\begin{figure}[tp]
    \includegraphics{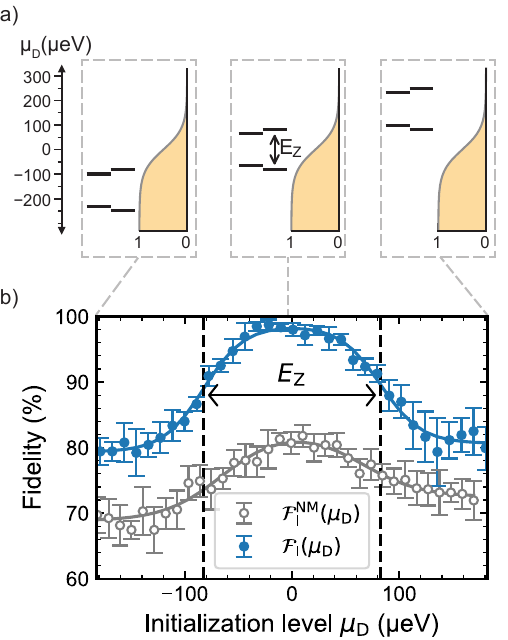}
    \caption{
    a) Cartoons of the electron spin states situated  about the Fermi reservoir in three different regimes: i) plunge ($\mu_{\rm D} \ll E_{\rm Z}$), where the donor is almost always loaded, ii) read ($\mu_{\rm D} \approx 0$), where the donor electron is able to tunnel to and from the reservoir and iii) empty ($\mu_{\rm D} \gg E_{\rm Z}$), where the donor electron is almost always unloaded. $E_{\rm Z}$ is the electron Zeeman energy.
    b) Total measurement fidelity as a function of donor potential $\mu_{\rm D}$. The grey data (open circle), $\mathcal{F}_{\rm I}^{\rm NM}(\mu_{\rm D})$, is the fidelity obtained with no monitoring, i.e. standard energy-dependent tunneling. The blue data (closed circle), $\mathcal{F}_{\rm I}(\mu_{\rm D})$, is obtained with real time monitoring by the Bayesian Maxwell's demon [$\mathcal{F}_{\rm I}(\mu_{\rm D})$, blue]. The lines are guides to the eye.
    The Maxwell's demon's intervention yields a drastically improved initialization fidelity over a wide range of donor potentials, making the effect robust against even large electrostatic detunings.
    }
    \label{fig:main_figure_5}
\end{figure}

Therefore, we fit the data to a modified version of Eq.~\ref{eq:bayesian_update_spin_down} to account for missed blips,
\begin{align}
    f_{\rm fit} &= \pgiv{\downarrow}{\neg \bm{B}^N} - P_M \nonumber \\
                    &= \Big[1 + \frac{1 - \prob{\downarrow}}{\prob{\downarrow}} e^{-t_{\rm obs}(\rate{\uparrow}{out} - \rate{\downarrow}{out})}  \Big]^{-1} - P_M.
    \label{eq:fit_function}    
\end{align}
While we estimate $P_M$ above, it is treated as a free fitting parameter.
The fit yields a total experiment fidelity approaching $\mathcal{F} = \SI{98.39(8)}{\percent}$ for $t_{\rm obs} > 10$~ms, which gives an initialization fidelity (by Eq.~\ref{eq:experiment_fidelity}) approaching $\mathcal{F}_{\rm I} = \SI{98.9(4)}{\percent}$. 
This is in close agreement with the limit predicted on the basis of the rise time limitation of the transimpedance amplifier, Eq.~\ref{eq:limit_infidelity_BW}. 
From this we deduce that a qubit initialization fidelity $\mathcal{F}_{\rm I} = 99.9$\% would be achieved if the amplifier chain had a cutoff frequency of \SI{300}{\kilo\hertz}, or if the electron tunnel-in rate is slowed down to \SI{880}{\second^{-1}}. In the absence of Maxwell's demon initialization, the bare fidelity is $\mathcal{F}_{\rm I}^{\rm NM}\approx 80$\% (Figure \ref{fig:main_figure_5}b), showing that our method reduces the initialization error by a factor 20.
The bare fidelity closely matches the prior $\ket{\downarrow}$ probability $\mathcal{F}_{\rm I}(0) \approx 78\%$ extracted from the fit.

Next we consider (and rule out) other potential error channels that could impose an upper bound to the initialization fidelity. A possible mechanism could be the spurious spin excitation caused by absorption of thermal phonons. However, for an electron spin in silicon in a magnetic field $B_0 > 1$~T, the spin excitation rate $W_{\downarrow\uparrow}$ is many orders of magnitude lower than the spin decay rate $W_{\uparrow\downarrow}$, which itself is about $3$ orders of magnitude lower \cite{Tenberg2019} than the tunnel rates. Since the observation time is much shorter than the relaxation and excitation rates (i.e.~$t_{\rm obs} \ll W_{\uparrow\downarrow}^{-1} \ll W_{\downarrow\uparrow}^{-1}$),
the number of thermally excited electrons recorded throughout this experiment is expected to be negligible. 

In the high-field limit $\gamma_{\rm e} B_0 \gg A$, the eigenstates of the electron-nuclear system are the simple tensor products states of the electron ($\ket{\downarrow},\ket{\uparrow}$) and nuclear ($\ket{\Downarrow},\ket{\Uparrow}$) basis states. We now consider the effect of the finite $B_0$ in mixing the electron-nuclear eigenstates as a possible explanation for the observed  -bound in $\ket{\downarrow}$ preparation fidelity. In this experiment, the device is operated in a magnetic field of $\SI{1.423}{\tesla}$ (see Appendix \ref{app:apparatus}), giving an electron Zeeman splitting of $\gamma_{\rm e} B_0 \approx \SI{39.8}{\giga\hertz}$. For an isotropic hyperfine interaction strength of $A \approx \SI{116}{\mega\hertz}$, this gives a deviation from the high-field limit electron spin eigenstates of only $10^{-5}$. In the context of Bayesian initialization, this is not the limiting factor in the present experiment.

The average time spent initializing the qubit $\bar{t}_{\rm init}$ is determined by the number of times the electron qubit is reset (i.e.~tunnels out and another tunnels back in) during observation.
For a given $t_{\rm obs}$, the total initialization time increases with the average electron reset time ($\tau_{\rm in} = \frac{1}{\rate{\uparrow}{in} + \rate{\downarrow}{in}}$) scaled by the likelihood of observing no blips during readout (the denominator in Eq.~\ref{eq:bayesian_update_full}),
\begin{equation}
    \bar{t}_{\rm init} = \frac{\tau_{\rm in}}{\likelihood{\downarrow}{\neg \bm{B}^N} \prob{\downarrow} + \likelihood{\uparrow}{\neg \bm{B}^N} \prob{\uparrow}}.
\end{equation}
For this experiment the average initialization time is $\bar{t}_{\rm init} \approx \SI{40}{\milli\second}$ with $t_{\rm obs} = \SI{10}{\milli\second}$.

Our initialization method compares favorably to alternatives that use additional quantum resources. Exploiting \gls*{qnd} readout, Yoneda \emph{et al.} achieve a single electron ground-state initialization fidelity of 95.9\% with a conventional majority vote of 20 consecutive reads  \cite{Yoneda2020}.
Extending beyond majority voting to supermajority voting (i.e. moving the majority threshold above 50\%), a $\ket{\downarrow}$ initialization fidelity of $\mathcal{F}_{\rm I} = 99.6\%$ was reached with at least a 76.6\% majority vote spin-down in 10 consecutive reads, on an approximately $\SI{600}{\micro\second}$ timescale.
Similarly, Philips \emph{et al.} \cite{philips2022universal} achieve up to 98\% single-electron ground-state initialization in a six qubit register with a 100\% majority vote of three consecutive \gls*{qnd} reads, on a timescale of about $\SI{300}{\micro\second}$.
Our method achieves 98.9(4)\% which is comparable to the state of the art while using fewer quantum resources. That the measurement time (approximately $\SI{40}{\milli\second}$) here is longer than in other methods should not be taken as a fundamental feature of our system. A device with faster tunnel rates could achieve similar fidelities in much shorter times. The generic scaling of initialization fidelity with observation time, expressed in multiples of the inverse tunnel rate, is shown in Fig.~\ref{fig:main_figure_model}b.


The above analysis is conducted by tuning the donor electrochemical potential $\mu_{\rm D}$ in alignment with the Fermi level. Next we investigate how the initialization fidelity varies with the initialization level, i.e.~the position of donor electrochemical potential $\mu_{\rm D}$ relative to $E_{\rm F}$ during the measurement phase. The donor initialization level is controlled by the virtual gate VG (see Figure \ref{fig:main_figure_1}a) which maintains a constant \gls*{set} potential.

In standard energy-dependent tunneling readout, the dependence of $\rate{\uparrow}{out}, \rate{\downarrow}{out}$ on $\mu_{\rm D}$ is reflected in a tuning dependence of the initialization fidelity (Fig.~\ref{fig:main_figure_5}b, grey). 
Conversely, the Bayesian Maxwell's demon proves to be very effective in reducing the tuning dependence, as shown by the plateau in $\mathcal{F}_{\rm I}$ (blue) extending for about half the range corresponding to the electron spin Zeeman energy.
The range over which the mean data remain within 3\% of the maximum value of each curve is approximately $40\%$ larger when Maxwell's demon is employed ($\SI{90}{\micro\electronvolt}$ cf.~$\SI{64}{\micro\electronvolt}$). Therefore, Maxwell's demon not only improves the initialization fidelity in absolute terms, but also makes it more robust against drift in the device tuning.

While common error-correcting codes place strict bounds on state preparation and measurement errors, they also typically require that the time taken for these steps is similar to the gate operation time. This defines the ``clock cycle'' of a quantum processor. In our case, the initialization and readout steps (about \SI{10}{\milli\second}) take significantly longer than the gate times (about \SI{100}{\nano\second}). We can slow down the gate operations by reducing the microwave power, however the coherence time of the system $T_2 \sim\SI{100}{\micro\second}$ then becomes a limiting factor. 
Improving initialization and readout times is thus a priority. 

For spin readout based on energy-dependent tunneling \cite{Elzerman2004,Morello2010}, decreasing initialization and readout times to \SI{100}{\nano\second} requires increasing the qubit-reservoir tunnel coupling to $\rate{\uparrow}{out} \approx 10^{7}~\mathrm{s}^{-1}$.
The ultimate speed limit for this method is set by the value of tunnel couplings where the electron starts to behave as a Kondo impurity \cite{Goldhaber1998,Cronenwett1998,Lansbergen2009}. At that point one loses the isolated nature of the spin qubit, whose levels hybridize with the reservoir states. The theoretical analysis of the impact of Kondo physics on readout speed is an ongoing project and will be discussed in future work.

Alternatively, fast spin readout can be achieved using PSB in two-electron systems \cite{Petta2005,barthel2009rapid}. This method works also when one of the two electrons is hosted by a donor coupled to a surface quantum dot. A hybrid donor-dot device has been operated with interdot tunnel couplings approximately $10^9~\mathrm{s}^{-1}$ in combination with latched electron readout \cite{HarveyCollard2018}. In combination with PSB, our method could be used to first prepare a fiducial one-electron $\ket{\downarrow}$ state, and then select the triplet or even-parity outcome of PSB readout to obtain a two-qubit register initialized in the $\ket{\downarrow\downarrow}$ state. 

\section{Electron temperature and spin-dependent tunneling}
\label{sec:electron_temp}

So far we have described the action of the Bayesian Maxwell's demon in terms of its impact on the fidelity $\mathcal{F}_{\rm I}$ of initializing the qubit in the $\ket{\downarrow}$ state. The increased fidelity upon real time observation by the demon should intuitively be linked to a reduction in the effective temperature of the system. In this section we attempt to quantify such a cooling effect. 

\begin{figure}[tbp]
    \centering
    \includegraphics[width=\linewidth]{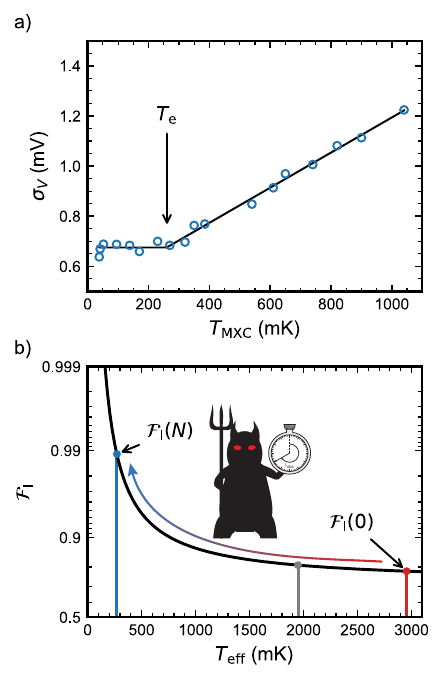}
    \caption{
    a) The electron temperature 
    $T_{\rm e} = \SI{260}{\milli\kelvin}$ is determined by the onset of an excess thermal broadening $\sigma_{\rm V}$ of the SET Coulomb peaks when increasing the temperature $T_{\rm MXC}$ of the dilution refrigerator's mixing chamber. 
    b) The initialization fidelity prior to the Maxwell demon's intervention (shown in red) is $\mathcal{F}_{\rm I}(0) = 78$\% which corresponds to the Fermi population at $T_{\rm eff} = \SI{2.95}{\kelvin}$.
    With Maxwell's demon monitoring, $\mathcal{F}_{\rm I}$ reaches approximately 99\% (shown in blue), which corresponds to an effective temperature $T_{\rm eff} = \SI{270}{\milli\kelvin}$.
    }
    \label{fig:main_figure_6}
\end{figure}

The starting point is the electron temperature $T_{\rm e}$ in the SET island. We measure $T_{\rm e} \approx \SI{260}{\milli\kelvin}$ using the standard method of observing the broadening on the SET Coulomb peaks \cite{beenakker1991theory,maradan2014gaas} as a function of the refrigerator mixing chamber temperature (Fig.~\ref{fig:main_figure_6}a). This value is higher than the refrigerator base temperature approximately $20$~mK due to the modest (10~dB) attenuation along the microwave line, necessary to allow for high-power delivery for nuclear spin control. Moreover, the charge reservoir we use for spin discrimination (the SET island) is also used as part of that charge sensing device, and thus subjected to a sporadic current flow which can affect its temperature. As we detail below, the attempt to relate $\mathcal{F}_{\rm I}$ to $T_{\rm e}$ 
highlights some long-standing (but seldom discussed) inconsistencies in the standard models of spin-dependent tunneling.

We can define an \emph{effective} temperature $T_{\rm eff}$ that determines $\mathcal{F}_{\rm I}$ from the tunnel rates between the donor and the Fermi reservoir. Let us call $H'$ the Hamiltonian describing the electron tunnel coupling between donor and SET island. Applying Fermi's golden rule, the (elastic) tunnel rate of a $\ket{\uparrow}$ electron from the reservoir to the empty donor can be written as:
\begin{equation}
    \rate{\uparrow}{in}(T_{\rm eff}) = \dfrac{2\pi}{\hbar} \left|\bra{0} H' \ket{\uparrow}\right|^2 n(E_{\uparrow}) f(E_\uparrow),
\end{equation}
where $\bra{0} H' \ket{\uparrow}$ is the tunneling transition matrix element between initial ($\ket{0}$, i.e. donor ionized) and final ($\ket{\uparrow}$ electron on the donor) state, $H'$ is the tunnel coupling Hamiltonian,
$n(E)$ is the reservoir density of states, and $E_{\uparrow} \approx \SI{82.5}{\micro\electronvolt}$ is the Zeeman energy of the $\ket{\uparrow}$ electron, having set $E_{\rm F}=0$ by convention. 
The dependence on $T_{\rm eff}$ is given by the Fermi function.
Similarly, the tunnel rate of a $\ket{\downarrow}$ electron to the reservoir is
\begin{equation}
    \rate{\downarrow}{in}(T_{\rm eff}) = \dfrac{2\pi}{\hbar} \left|\bra{0} H' \ket{\downarrow}\right|^2 n(E_{\downarrow})f(E_\downarrow).
\end{equation}
At the ideal readout position, $E_\downarrow = - E_\uparrow$.

Here, and in the near totality of the literature on spin-dependent tunneling in quantum dots, two implicit assumptions are made: (i) that $\bra{\downarrow} H' \ket{0} = \bra{\uparrow} H' \ket{0}$, i.e. the tunnel barrier seen by either spin state is the same, and (ii) that the density of states is constant over the range of interest, i.e. $n(E_{\uparrow}) = n(E_{\downarrow})$. 
Assumption (i) is physically justified on the basis that the barrier profile is, in fact, the edge of the conduction band (Figure \ref{fig:main_figure_1}b), which itself is subjected (to a very good approximation) to the same Zeeman splitting as the donor-bound electron \cite{Amasha2008}.
Assumption (ii) is carried over from the physics of spins in quantum dots tunnel coupled to a two-dimensional electron gas, where $n(E)= $~const. \cite{ando1982electronic}.  
We know that our situation is rather different, because the charge reservoir into which the electron tunnels is an SET island with about $100$~electrons, i.e.~a near-zero dimensional confining potential. 
A simple estimate yields a single-particle level spacing $\approx \SI{24}{\micro\electronvolt}$\cite{Morello2009}, smaller than the Zeeman splitting and comparable to the temperature of the experiment.

In the absence of the Maxwell's demon, the expected initialization fidelity is simply:
\begin{equation}
    \mathcal{F}_{\rm I}^{\rm NM} = \frac{\rate{\downarrow}{in}}{\rate{\downarrow}{in} + \rate{\uparrow}{in}}. \label{eq:loading_tunnel_rates}
\end{equation}
This model fails to reproduce the no-monitoring (grey) data in Fig.~\ref{fig:main_figure_5}b. At very negative initialization level, where $f(E_\uparrow) \approx f(E_\downarrow) \approx 1$ and $\rate{\uparrow}{in} \approx \rate{\downarrow}{in}$, the model predicts $\mathcal{F}_{\rm I}^{\rm NM}\approx 50$\%, versus the observed $\approx 70$\%. Such behavior is ubiquitous in electron and hole spin qubits in GaAs \cite{Elzerman2004}, Si \cite{Morello2010,spence2022spin} and Ge \cite{vukusic2018}. The initial spin-up fraction upon loading a random electron is seen in spin-lattice relaxation ($T_1$) experiments, and is usually around 30\%--40\% instead of the expected 50\%. We are not aware of papers reporting a 50\% spin-up fraction from random loading.  

More detailed analyses of spin-dependent tunnel rates in quantum dots \cite{Amasha2008,house2013detection} further confirm such discrepancies. Following Ref.~\cite{Amasha2008}, we introduce a phenomenological parameter $\chi$ as:
\begin{equation}
    \chi = \frac{n(E_{\uparrow}) \left|\bra{\uparrow} H' \ket{0}\right|^2 }{n(E_{\downarrow}) \left|\bra{\downarrow} H' \ket{0}\right|^2 }.
\end{equation}
We allow $\chi \neq 1$, but assume that it does not depend on $T_{\rm eff}$. Defining the ratio of tunnel-in rates:
\begin{equation}
    R_{\rm in}(T_{\rm eff}) = \dfrac{\rate{\uparrow}{in}}{\rate{\downarrow}{in}} = \chi \dfrac{f(E_\uparrow)}{f(E_\downarrow)},
\end{equation}
the bare initialization fidelity becomes:
\begin{equation}
\mathcal{F}_{\rm I}^{\rm NM} = \frac{1}{1 + R_{\rm in}} = \frac{1}{1 + \chi \frac{f(E_\uparrow)}{f(E_\downarrow)}}.\label{eq:loading_fidelity}
\end{equation}

The no-monitoring data at very negative initialization level in Fig.~\ref{fig:main_figure_5}b can be used as a proxy for $\mathcal{F}_{\rm I}^{\rm NM}$ when $T_{\rm eff}\rightarrow \infty$, because it describes the case where $f(E_\uparrow)=f(E_\downarrow)$. From $\mathcal{F}_{\rm I}^{\rm NM}(\mu_{\rm D}\ll 0)=72\%$ at deep plunge we extract $\chi=0.388$. The solid line in Fig.~\ref{fig:main_figure_6}b describes $\mathcal{F}_{\rm I}^{\rm NM}$ as a function of $T_{\rm eff}$, calculated from Eq.~\ref{eq:loading_fidelity}. From Fig.~\ref{fig:main_figure_5}b, the no-monitoring data at the optimal readout point yield $\mathcal{F}_{\rm I}^{\rm NM}\approx 81\%$, which corresponds to $T_{\rm eff}\approx 2$~K (grey dot). From fitting the data in Fig.~\ref{fig:main_figure_4} we obtain $\mathcal{F}_{\rm I}(N)\approx 78\%$ for $N=0$, which corresponds to $T_{\rm eff}\approx 3$~K (red dot). The action of the Bayesian Maxwell's demon resulted in $\mathcal{F}_{\rm I}(N)\approx 99\%$ for $N=2000$. To this result, following the black line in Fig.~\ref{fig:main_figure_6}b, we can attribute an effective temperature $T_{\rm eff}(N)\approx 270$~mK, indicating that the demon has reduced the effective temperature by about an order of magnitude.

The above discussion does allow us to frame the action of the Bayesian Maxwell's demon as a form of cooling. However, it also highlights the fundamental inadequacy of simple spin-dependent tunneling models to quantitatively describe the data.

\section{Conclusions} 
We have presented a simple and effective method to drastically improve the initialization fidelity of an electron spin qubit. Starting from a poorly thermalized charge reservoir, we have achieved $\mathcal{F}_{\rm I}=98.9(4)$\%. The method can be described as a form of Bayesian Maxwell's demon, that updates its confidence about the true state of the qubit while performing a negative-result measurement, i.e.~watching the absence of a tunnel-out event. The effect can be conceptually described as drawing the electron from a colder reservoir, but a quantitative analysis reveals that a simple spin-dependent tunneling model is not adequate to capture the details.

Our method is purely based on classical software and does not require additional quantum resources such as ancilla qubits. We used a nuclear spin ancilla solely to verify the initialization fidelity beyond the intrinsic electron spin readout fidelity, but once the method is ``trusted'', it can be applied to single qubits without ancillas. Further increasing $\mathcal{F}_{\rm I}$ to 99.9\% will be possible by adopting faster readout techniques, e.g. with cold baseband \cite{tracy2016single} or radio-frequency \cite{barthel2009rapid,keith2019single} amplifiers, or in cavity-based setups \cite{Petersson2012,Mi2018,Zheng2019,Borjans2021}. 

Future work will focus on integrating this method within multi-qubit systems \cite{madzik2022precision,madzik2021conditional}. We expect it will improve the fidelity in preparing highly entangled quantum states, and enable sufficient state preparation fidelity to implement quantum error correction codes \cite{Fowler2012,Andrist2011} at a fault-tolerant level.  

\section*{Acknowledgements}
We thank N. Ares and S. Asaad for discussions. 
The research was supported by the Australian Research Council (Grant No. CE170100012), the US Army Research Office (Contract no. W911NF-17-1-0200), and the Australian Department of Industry, Innovation and Science (Grant No. AUSMURI000002). 
M.~A.~I.~J. acknowledges the support of an Australian Government Research Training Program Scholarship. 
We acknowledge support from the Australian National Fabrication Facility (ANFF). All statements of fact, opinion or conclusions contained herein are those of the authors and should not be construed as representing the official views or policies of the U.S. Government.

\appendix
\section{Device fabrication and experimental apparatus}
\label{app:apparatus}
The device is fabricated on a \SI{0.9}{\micro\metre} thick enriched $^{28}$Si epilayer with a residual $^{29}$Si concentration of  \SI{730}{ppm}.
The epilayer is implanted with $^{31}$P ions at an energy of \SI{10}{\kilo\electronvolt} and fluence of \SI{1.4e12}{\centi\metre^{-2}}. 
Further details on the device design, fabrication and ion implantation are given in Ref.~\cite{madzik2022precision}.

The device is mounted in a copper enclosure and wire-bonded to a gold-plated printed circuit board with aluminum wires.
The sample is then mounted in a Bluefors LD400 cryogen-free dilution refrigerator with a base temperature of \SI{22}{\milli\kelvin}.
The sample is mounted in the centre of a superconducting solenoid which is set to a magnetic field of approximately $\SI{1.423}{\tesla}$.

The electronic apparatus used to operate the sample is identical to that in Ref.~\cite{Asaad2020}.

The Maxwell's demon is implemented in the \gls*{fpga} hardware of a Keysight M3300A digitizer.
The Maxwell's demon takes the form of a simple state machine with three states: observation, trigger and post-trigger wait.
In the observation state, a hardware counter increments while a blip is not detected ($\neg B_n$) and resets otherwise ($B_n$).
Once the counter value $N$ reaches the prescribed observation time $N \tsample = t_{\rm obs}$, the demon's state changes to trigger.
While in the trigger state, an output trigger is asserted for a fixed duration, and then the state automatically changes to the post-trigger wait which waits for the measurement pulse sequence to complete before automatically transitioning back to the observation state.

The logic is designed to operate in a single cycle of the onboard clock to minimize the response time of the hardware.
The onboard clock runs at \SI{100}{\mega\hertz}.
There is a single cycle delay between $t_{\rm obs}$ being met and the trigger being asserted, which accounts for an approximately $\SI{100}{\nano\second}$ latency.

We note that the \gls*{set} signal is downsampled to a rate of $f_{\rm s} = \SI{100}{\kilo S\per\second}$ ($\tsample = \SI{10}{\micro\second}$), or two times the bandwidth $f_{\rm s} = 2 f_{\rm c}$.
The downsampling is a decimation from \SI{100}{\mega S\per\second}  without internal low-pass filtering and is performed by the proprietary Keysight digital acquisition system.
As $t_{\rm rise} < \tsample$, a higher proportion of missed blips may occur, which could result in a worse initialization fidelity.

\section{Iterative Bayesian update model}
\label{app:bayesian}
The probability of having a $\ket{\downarrow}$ electron after monitoring the sensor current with a digital system can be described by an iterative Bayesian update process.
A single sample, $B_n$, is acquired at times $n \cdot \tsample$ for positive integers $n$.
If no blip is detected in the first measured sample ($\neg B_1$), i.e.~the signal threshold is not exceeded, the $\ket{\downarrow}$ probability is updated based on the likelihood of the electron \emph{not} tunneling for either state ($\likelihood{\downarrow}{\neg B_1}$ and $\likelihood{\uparrow}{\neg B_1}$).
\begin{align}
    \label{eq:Bayes_rule}
    \pgiv{\downarrow}{\neg B_1} &= \dfrac{\likelihood{\downarrow}{\neg B_1} \prob{\downarrow}}{\sum_\psi \likelihood{\psi}{\neg B_1}\prob{\psi}} \\
        &= \dfrac{\likelihood{\downarrow}{\neg B_1} \prob{\downarrow}}
            {\likelihood{\downarrow}{\neg B_1} \prob{\downarrow} + \likelihood{\uparrow}{\neg B_1} \prob{\uparrow}}
\end{align}
where $\prob{\downarrow}$ is the $\ket{\downarrow}$ probability prior to obtaining a sample, with $\prob{\downarrow} + \prob{\uparrow} = 1.$
The denominator is the marginal likelihood function over each spin state ${\psi \in \{\downarrow, \uparrow\}}$.
The likelihood of not observing a blip during a single sample is
\begin{align}
    \likelihood{\uparrow}{\neg B_1} &= e^{-\tsample \rate{\uparrow}{out}}\label{eq:likelihood_up}  \\
    \likelihood{\downarrow}{\neg B_1} &= e^{-\tsample \rate{\downarrow}{out}} \label{eq:likelihood_down},
\end{align}
    for a $\ket{\uparrow}$ and $\ket{\downarrow}$ electron, respectively.
    When the next sample, $B_2$ comes in, we can repeat the process as in Eq.~\eqref{eq:Bayes_rule} but we replace the prior with the first posterior, $\pgiv{\downarrow}{\neg B_1}$.
\begin{align}
    \begin{split}
    \pgiv{\downarrow}{\neg B_2, \neg B_1}  &= \dfrac{\likelihood{\downarrow}{\neg B_2} }{\sum_\psi \likelihood{\psi}{\neg B_2}\prob{\psi}} \\
    & \times \dfrac{\likelihood{\downarrow}{\neg B_1} \prob{\downarrow}}{\sum_\psi \likelihood{\psi}{\neg B_1}\prob{\psi}}
    \end{split} \\ 
    &= \dfrac{\likelihood{\downarrow}{\neg \bm{B}^2} \prob{\downarrow}}{\sum_\psi \likelihood{\psi}{\neg \bm{B}^2}\prob{\psi}},
    \label{eq:two_samples_update}
\end{align}
where $\bm{B}^2 = (B_2, B_1)$ represents the collection of $N = 2$ samples and the likelihood functions of two samples is simply the product of the likelihood function of each sample.
Substituting Eqs.~\eqref{eq:likelihood_down} and \eqref{eq:likelihood_up} into Eq.~\eqref{eq:two_samples_update} gives
\begin{equation}
    \pgiv{\downarrow}{\neg B_2, \neg B_1} = \frac{e^{-2 \tsample \rate{\downarrow}{out}} \prob{\downarrow}}
            {e^{-2 \tsample \rate{\downarrow}{out}} \prob{\downarrow} + e^{-2\tsample \rate{\uparrow}{out}} \prob{\uparrow}}
\end{equation}

When the data are independent we can reformulate this sequential process as a one-step process\cite{Sivia2006}, collecting the $N$ samples and applying them `all at once'.
The data are independent in the case of tunneling because an object has the same likelihood of tunneling at all times over a fixed duration, i.e.~$\likelihood{\downarrow}{\neg B_n} = \likelihood{\downarrow}{\neg B_{n+1}}$ for all $n$, and likewise for the $\ket{\uparrow}$ case.

\section{Accounting for imperfect detection}
\label{app:missing_blips}

A missed blip is the result of an electron tunneling to the \gls*{set} island followed by an electron rejoining the donor \emph{before} the amplifier output signal rises above the detection threshold $S_{\rm th}$.
Since these events are fundamentally undetectable with the measurement apparatus, we do not know when they occur.
During an initialization trace where no samples exceed the detection threshold, it is possible that a blip is missed at any point in the trace.
Moreover, it is also possible (though unlikely) that multiple blips are missed in a single trace.
This makes the final $\ket{\downarrow}$ probability difficult to exactly calculate since the different times at which the fast tunneling events occur yield different outcomes.
Despite this inherent complexity, the initialization fidelity from a missed blip has a lower bound set by the prior $\ket{\downarrow}$ probability $P(\downarrow) \approx 78\%$.
We now formally extend the model presented in the Sec.~\ref{sec:bayesian_maxwell_demon} and Appendix \ref{app:bayesian} to account for unobserved tunneling events.

The intuitive motivation for the following treatment is that the final initialization fidelity arises from two distinct cases: initialization where a blip is not missed $\pgiv{\downarrow}{\neg \bm{B}^N , \neg M}$ and where a blip is missed $\pgiv{\downarrow}{\neg \bm{B}^N , M}$.
A missed blip consequently reduces the average initialization fidelity. 
Since the probability of missing a blip is inherently rare (Eq.~\ref{eq:limit_infidelity_BW}), this yields only a small difference in the final outcome.
The probability of having a $\ket{\downarrow}$ electron after readout is therefore
\begin{equation}
    \begin{split}
    \pgiv{\downarrow}{\neg \bm{B}^N} = & \pgiv{\downarrow}{\neg \bm{B}^N , \neg M} (1 - P_M) \\&+ \pgiv{\downarrow}{\neg \bm{B}^N , M} P_M.
    \end{split}\label{eq:missed_blip}
\end{equation}
Since the unobserved tunneling events happen at an unknown time, the conditional probability \pgiv{\downarrow}{\neg \bm{B}^N , M} is difficult to define.
However, we know the $\ket{\downarrow}$ probability is lower in the cases where the electron does not tunnel due to the effect of the Maxwell's demon, i.e.~
\begin{equation}
    \pgiv{\downarrow}{\neg \bm{B}^N , M} < \pgiv{\downarrow}{\neg \bm{B}^N , \neg M}.
\end{equation}
We can then reformulate Eq.~\ref{eq:missed_blip} as
\begin{align}
\begin{split}
    \pgiv{\downarrow}{\neg \bm{B}^N} = & \pgiv{\downarrow}{\neg \bm{B}^N , \neg M}\\
    &- P_M [\pgiv{\downarrow}{\neg \bm{B}^N , \neg M} - \pgiv{\downarrow}{\neg \bm{B}^N , M}]
    \end{split}\\
    &= \pgiv{\downarrow}{\neg \bm{B}^N , \neg M} - Z P_M,
\end{align}
where $Z = \pgiv{\downarrow}{\neg \bm{B}^N , \neg M} - \pgiv{\downarrow}{\neg \bm{B}^N , M}$ is a number between zero and $P(\downarrow)$ depending on when the electron rejoins the donor in time.
$Z$ takes on its maximum value when a new electron rejoins the donor in the final sample of the initialization trace.

Since $Z < 1$, the upper bound to initialization fidelity $\mathcal{F}_{\rm I}$ is $1 - Z P_M$ at $t_{\rm obs} \to \infty$.
This is higher than the conservative estimate $1 - P_M$ presented in the main text (Eq. \ref{eq:fit_function}).

\section{Maxwell's demon as a variable-strength observer}
\label{app:weak_measurement}
In this paper we present a Maxwell's demon that initializes a $\ket{\downarrow}$ electron using readout.
For long observation periods without a tunneling event, the initialization fidelity drastically improves as the demon performs a strong projective measurement onto the state $\ket{\downarrow}$.
Conversely, for short observation periods, the initialization fidelity improves moderately, according to the amount of information extracted from the system\cite{Korotkov2003}.
In this way, the Maxwell's demon is performing a weak measurement for $t_{\rm obs} \lesssim \rate{\uparrow}{out}^{-1}$.
Note that the measurement is weak only if no tunneling event occurs during $t_{\rm obs}$, since otherwise an electron tunneling projects the spin state to $\ket{\uparrow}$.

The information extracted from the system is described by a discrete-time Bayesian update process (Sec.~\ref{sec:bayesian_maxwell_demon} and Appendix \ref{app:bayesian}), but it can also be described by a continuous-time process.
We define the measurement strength under continuous readout as the projection onto the $\ket{\downarrow}$ state
\begin{equation}
        m(t_{\rm obs}) = \Tr{\rho(t_{\rm obs})   \ket{\downarrow}\!\bra{\downarrow}}
        \label{eq::measurement_strength_def}
\end{equation}
We now introduce a quantum master equation to describe the evolution of the electron state $\rho(t_{\rm obs})$ under continuous measurement.

At the readout position, the donor electron can either be loaded $(\ket{\uparrow}$ or $\ket{\downarrow})$ or unloaded $ (\ket{0})$.
These three states now form our basis $\{ \ket{\uparrow},\ket{\downarrow},\ket{0}\}$.
The electron state evolves according to
\begin{equation}
    \frac{d}{dt}\rho(t) = L \rho(t),
    \label{eq:master_equation}
\end{equation}
with Liouvillian\cite{Emary2012}
\begin{equation}
    L = \begin{pmatrix} 
        -W_{\uparrow\downarrow} -\rate{\uparrow}{out} & W_{\downarrow\uparrow}   & \rate{\uparrow}{in} \\
        W_{\uparrow\downarrow} & -\rate{\downarrow}{out} -W_{\downarrow\uparrow} & \rate{\downarrow}{in} \\
        \rate{\uparrow}{out} & \rate{\downarrow}{out} & -\rate{\uparrow}{in} - \rate{\downarrow}{in}
    \end{pmatrix},
    \label{eq:liouvillian}
\end{equation}
and where $W_{\uparrow\downarrow(\downarrow\uparrow)}$ are the electron spin-phonon relaxation (excitation) rates.
For the present consideration, the electron phonon relaxation and excitation rates $W_{\downarrow\uparrow} \ll W_{\uparrow\downarrow} < T_1^{-1} \approx \SI{1}{\second^{-1}}$\cite{Tenberg2019} are much smaller than the tunnel rates, and so are treated as zero, giving
\begin{equation}
    L = \begin{pmatrix} 
                        -\rate{\uparrow}{out} & 0 & \rate{\uparrow}{in} \\
                        0 & -\rate{\downarrow}{out} & \rate{\downarrow}{in} \\
                        \rate{\uparrow}{out} & \rate{\downarrow}{out} & -\rate{\uparrow}{in} - \rate{\downarrow}{in}
                    \end{pmatrix}.
\end{equation}

We assume an initial, randomly prepared electron state ${\ket{\downarrow}}$ with probability $\prob{\downarrow}$ and $\ket{\uparrow}$ with probability ${\prob{\uparrow}} = 1- \prob{\downarrow}$,
\begin{equation}
\rho(t = 0) = \begin{pmatrix}
                    1 -\prob{\downarrow} \\
                    \prob{\downarrow} \\
                    0
                  \end{pmatrix}.
\end{equation}
By constraining $\rho_{3}(t_{\rm obs}) = 0$, we describe the evolution $\rho(t_{\rm obs})$ in the absence of any observed electron tunneling events.
The electron spin state then evolves as
\begin{equation}
\rho(t_{\rm obs}) = 
    C
    \begin{pmatrix}
        (1-\prob{\downarrow}) e^{-\rate{\uparrow}{out} t_{\rm obs}} \\
        \prob{\downarrow} e^{-\rate{\downarrow}{out} t_{\rm obs}}\\
        0
    \end{pmatrix},
    \label{eq::post_measurement_state}
\end{equation}
where 
\begin{equation}
    C = \dfrac{1}{(1-\prob{\downarrow})e^{-\rate{\uparrow}{out} t_{\rm obs}} + \prob{\downarrow} e^{-\rate{\downarrow}{out} t_{\rm obs}}}   
\end{equation}
normalizes the vector at each time $t_{\rm obs}$.

The measurement strength then becomes
\begin{align}
        m(t_{\rm obs}) &= \rho_{2}(t_{\rm obs}) \\
        &= \frac{\prob{\downarrow} e^{-\rate{\downarrow}{out} t_{\rm obs}}}
          {\prob{\downarrow} e^{-\rate{\downarrow}{out} t_{\rm obs}} + (1-\prob{\downarrow}) e^{-\rate{\uparrow}{out} t_{\rm obs}}},
          \label{eq::measurement_strength}
\end{align}
which agrees with the discrete-time Bayesian update approach detailed in the main text (Sec.~\ref{sec:bayesian_maxwell_demon}, Eq.~\ref{eq:bayesian_update_spin_down}). 

The rate of change of the measurement strength, $\frac{\partial}{\partial t_{\rm obs}} m(t_{\rm obs})$, can be understood as the rate at which we accumulate knowledge of the spin state.
Note that the rate of information gain depends on the initial population $\prob{\downarrow}$.
This can be intuitively explained by considering the extreme cases of $\prob{\downarrow}=1$ or 0.
Here, we know with certainty that the electron is initially $\ket{\downarrow}$ or $\ket{\uparrow}$.
Since there is no knowledge to be gained, $\frac{\partial}{\partial t_{\rm obs}} m(t_{\rm obs})$ = 0.

In the limit of measuring for infinite time without observing a tunneling event, the measurement strength (and the final electron $\ket{\downarrow}$ probability) approaches unity:
\begin{equation}
    \lim_{t_{\rm obs}\to \infty} m(t_{\rm obs}) = 1.
\end{equation}

\section{Nuclear control fidelity}
\label{app:control_fidelity}

The control fidelity in these experiments is the accuracy of the \gls*{nmr} $\pi$ pulse used to map the electron spin state to the nucleus.
The fidelity of an \gls*{nmr} $\pi$ pulse can be hampered by both a frequency mismatch of the applied stimulus and the system resonance ($\epsilon$) and an over- or under-rotation caused by miscalibrated pulses.
We define the control fidelity based on the Rabi formula
\begin{equation}
    \mathcal{F}_{\rm C} = \frac{\Delta^2}{\Delta^2 + \epsilon^2} \sin^2\left(\frac{\Omega}{2} t_{\rm pulse}\right),
    \label{eq:control_fidelity}
\end{equation}
where $\Delta$ is the drive strength, $\Omega = \sqrt{\Delta^2 + \epsilon^2}$ is the Rabi frequency and $t_{\rm pulse}$ is the duration of the pulse.

A nonzero frequency detuning $\epsilon$ reduces the amplitude of the oscillation while an imperfect pulse duration $t_{\rm pulse}$ leads to over- or under-rotation of the spin.
Prior to each experiment we perform calibration routines that measure the \gls*{nmr} frequency corresponding to the electron $\ket{\downarrow}$ state using Ramsey fringes to tune to within \SI{100}{\hertz}.
With the precision of $\epsilon \le \SI{100}{\hertz}$ the amplitude of the Rabi oscillations is expected to be 99.98\% with a measured frequency of $\Omega_R \approx \SI{7.6}{\kilo\hertz}$.

The spin rotation angle $\theta = \Omega t_{\rm pulse}$ can be determined by comparing \acrlong*{cp} and \acrlong*{cpmg} experiments, which allows us to extract the amount of over- or under-rotation of the spin\cite{Morton2005}. 
Note that this method is insensitive to the sign of the angle of error due to symmetry of $\sin^2(\theta/2)$ about $\theta = \pi$.
We find a rotation error $\sigma = \theta - \pi$ of $\sigma = \pm 0.143(7)$ rad with $t_{\rm pulse} = \SI{67}{\micro\second}$.

Substituting these values into Eq.~\ref{eq:control_fidelity} gives a final control fidelity of $\mathcal{F}_{\rm C} = 99.5(3)\%$.

\section{Nuclear readout fidelity}
\label{app:readout_fidelity}

The readout fidelity $\mathcal{F}_{\rm R}$ is the accuracy of correctly determining the nuclear spin state.
Since we perform repeated \gls*{qnd} readout of the nuclear spin \cite{Pla2013}, the readout fidelity is sensitive to both the accuracy with which the nuclear spin state can be determined, $\mathcal{F}_{\rm det}$, and how often the readout method causes the nuclear spin state to change during measurement, $\mathcal{F}_{\rm QND}$.
The nuclear readout fidelity can then be expressed as  
\begin{equation}
    \mathcal{F}_{\rm R} = \mathcal{F}_{\rm det} \cdot \mathcal{F}_{\rm QND}.
\end{equation}

\begin{figure}[htbp]
    \centering
    \includegraphics[width=\linewidth]{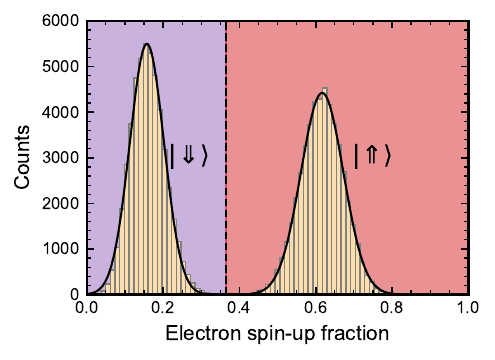}
    \caption{
    Electron $\ket{\uparrow}$ fraction recorded throughout experiment from \num{100000} nuclear reads which show two clearly resolved distributions.
    }
    \label{fig:readout_visibility}
\end{figure}

To determine the nuclear spin state we load an electron, perform an adiabatic \gls*{esr} pulse conditioned on the nucleus being $\ket{\Uparrow}$, and then read out the electron.
We repeat this process $n = 65$ times, and record the fraction of these electron reads that are $\ket{\uparrow}$.
Figure \ref{fig:readout_visibility} shows a histogram of the electron $\ket{\uparrow}$ fraction recorded from 100000 single nuclear spin readout shots.
The histogram shows two clearly resolved peaks which correspond to the nuclear $\ket{\Downarrow}$ and $\ket{\Uparrow}$ states.
We designate any electron $\ket{\uparrow}$ fraction greater than a chosen threshold (indicated by the dotted line) as resulting from a nuclear $\ket{\uparrow}$ (red region); otherwise it is designated a nuclear $\ket{\downarrow}$ result (purple region).

The visibility $V$ is a measure of how distinct the two peaks are, and is determined by fitting two Gaussian profiles to the peaks and calculating their overlap. 
The nuclear spin visibility provides a lower bound for the nuclear spin determination fidelity:
\begin{equation}
    V = \mathcal{F_\Downarrow} + \mathcal{F_\Uparrow} - 1 < \mathcal{F}_{\rm det}.
\end{equation}
The two Gaussian fits have a numerically calculated overlap of less than \num{5e-6} which gives a visibility of $V = 99.9995\%$, hence $\mathcal{F}_{\rm det} > 99.9995\%$.

While \gls*{qnd} readout allows for the nucleus to be repeatedly measured, \gls*{qnd} readout itself is imperfect due to ``ionization shock'' which can flip the nuclear spin when loading or unloading an electron.
This is due to the components of the electron-nuclear hyperfine interaction tensor $\underline{\underline{A}}$ that do not commute with the measurement operator, and thus violate the QND condition \cite{Braginsky1980}.
We find experimentally that each tunneling event has an average probability $p = \num{1.4e-6}$ of changing the nuclear spin state.
The probability of encountering zero erroneous spin flips during readout depends on the number of electron tunneling events that occur during readout which is on average the number of electron shots $n$,
\begin{equation}
    \mathcal{F}_{\rm QND} = (1 - p)^{n}.
\end{equation}
For a single nuclear readout, in which we perform $n = 65$ electron shots the average \gls*{qnd} fidelity is $\mathcal{F}_{\rm QND} = 99.99\%$.
Therefore, the average readout fidelity is $\mathcal{F}_{\rm R} = 99.99\%$.

\end{document}

%% file: preamble.tex
\usepackage{dcolumn}
\usepackage{bm}
\usepackage{hyperref}

\usepackage[utf8]{inputenc}

\usepackage{graphicx}

\usepackage{amsmath}
\usepackage{amssymb}
\usepackage{lipsum}
\usepackage{glossaries}

\usepackage{siunitx}
\usepackage{braket}

\usepackage{mathtools}
\DeclarePairedDelimiter{\Tr}{\textnormal{Tr}[}{]}

\newcommand{\prob}[1]{\ensuremath{P(#1)}}
\newcommand{\pgiv}[2]{\ensuremath{P(#1 \,\mid #2)}}
\newcommand{\likelihood}[2]{\ensuremath{\mathcal{L}(#1\mid #2)}}

\usepackage{natbib}

\usepackage{xcolor}

\definecolor{commentgreen}{RGB}{2,112,10}
\definecolor{eminence}{RGB}{108,48,130}
\definecolor{weborange}{RGB}{255,165,0}
\definecolor{frenchplum}{RGB}{129,20,83}
\usepackage{listings}

%% file: glossary.tex
\newacronym{set}{SET}{single electron transistor}
\newacronym{nmr}{NMR}{nuclear magnetic resonance}
\newacronym{esr}{ESR}{electron spin resonance}
\newacronym{spam}{SPAM}{state preparation and measurement}
\newacronym{qnd}{QND}{quantum nondemolition}
\newacronym{fpga}{FPGA}{field-progammable gate array}
\newacronym{qpc}{QPC}{quantum point contact}
\newacronym{cp}{CP}{Carr-Purcell}
\newacronym{cpmg}{CPMG}{Carr-Purcell-Meiboom-Gill}